\documentclass[a4paper,12pt]{article}
\usepackage{german,a4}[1998/07/08]
\usepackage{graphicx}
\begin{document}

\title{Lichtquanten und Molek"ule: Ein Beitrag zum Annus mirablis
\footnote{Johannes-Kepler-Vorlesung 2005, T"ubingen, 29. Juni.}}

\author{Norbert Straumann, Institut f\"ur Theoretische Physik, \\
  Universit\"at Z\"urich, Winterthurerstrasse 190, 8057 Z\"urich}
\date{}
\maketitle

\section{Einleitung}

\noindent Meine Damen und Herren,

\vspace{.5cm}

\noindent Ich f\"uhle mich sehr geehrt, diese \emph{Johannes-Kepler-Vorlesung} halten zu
d\"urfen. Gleichzeitig habe ich ein etwas beklemmendes Gef\"uhl, denn meine Aufgabe ist
nicht leicht.

"Uberall auf der Welt werden in diesem Jahr vor allem drei der f"unf Arbeiten des
denkw"urdigen Jahres 1905 gew"urdigt, die Einstein in den Monaten M"arz, Mai und Juni
verfasste und mit denen er als Physiker sein Jahrhundert "uberragt. Aber schon vor diesem
Jahr der Wunder hat Einstein wichtige Arbeiten "uber die molekulare Basis der Thermodynamik
in den ``\textit{Annalen der Physik}'' publiziert. In drei Abhandlungen entwickelte er
unabh"angig vom grossen Willard Gibbs die Grundlagen der Statistischen Mechanik. Man kann
nur staunen, mit welcher Sicherheit der 23j"ahrige diese prinzipiellen Probleme angeht.
Besonders fruchtbar sollten sp"ater seine hier gewonnenen Einsichten in Natur und Gr"osse
von \emph{Schwankungserscheinungen} werden, ohne die insbesondere Einstein's
revolution"arster Beitrag zur Physik -- seine Lichtquantenhypothese -- kaum denkbar w"are.
Dabei erkannte Einstein die wichtige Rolle, welche die sp"ater nach Boltzmann benannte
Konstante $k$, bzw. die damit verkn"upfte Avogadro'sche Zahl $N_A$ spielt. Als Einstein
diese Arbeiten verfasste, war ihm das kurz davor erschienene epochale Werk
``\textit{Elementary Principles of Statistical Mechanics}'' von Gibbs unbekannt. Aus diesem
Grund sind die drei Arbeiten von Einstein nicht als Marksteine in die Physik eingegangen,
aber f"ur ihn spielten sie in den n"achsten zwanzig Jahren wiederholt eine zentrale Rolle.
Mit den ersten drei Arbeiten von 1905 war Einstein im Besitz von drei unabh"angigen
Methoden zur Bestimmung der Avogadro-Zahl, und weitere sollten folgen.

Zu dieser Thematik ist k"urzlich ein sch"ones handschriftliches Manuskript von Einstein aus
dem Jahre 1910 aufgetaucht, als der umfangreiche Zangger-Nachlass endlich zug"anglich
wurde. Es handelt sich hier, wie Walter Hunziker und ich nach einem Hinweis von Robert
Schulmann rasch herausfanden, um einen s"auberlich aufgeschrieben Vortrag, den Einstein am
2. November 1910 vor der Physikalischen Gesellschaft Z"urich gehalten hat. Dieser tr"agt
den Titel ``\textit{"Uber das Boltzmann'sche Prinzip und einige unmittelbar aus demselben
fliessender Folgerungen}''. In diesem Referat entwickelt Einstein nach einigen allgemeinen
Bemerkungen "uber Thermodynamik und Statistische Mechanik Beispiele seiner wichtigsten
Einsichten, die er auf der Basis von Schwankungserscheinungen in den f"unf vorangegangenen
Jahren gewonnen hatte. Er diskutiert insbesondere die Implikationen "uber die ``Struktur
der W"armestrahlung'' und die Opaleszenz (optische Tr"ubung) des Lichtes aufgrund von
Dichteschwankungen.-- Ich werde auf diese Themen sp"ater eingehen.

Einstein beschliesst sein Referat mit einem Glaubensbekenntnis an eine umfassende
Kausalit"at:
\begin{quote}
``\textit{Indessen f"uhren uns die allenthalben sich bew"ahrenden Mittelwertgesetze sowie
die in jenen Gebieten feinster Wirkungen g"ultigen statistischen Gesetze "uber die
Schwankungen zu der "Uberzeugung, dass wir an der Voraussetzung der vollst"andigen kausalen
Verkn"upfung des Geschehens in der Theorie festhalten m"ussen, auch wenn wir nicht hoffen
d"urfen, durch verfeinerte Beobachtungen von der Natur die unmittelbare Best"atigung dieser
Auffassung je zu erlangen.''}
\end{quote}

Zu dieser Veranstaltung existiert ein Protokoll, in welchem am Schluss folgendes steht:
``\textit{Der Vortrag gibt zu einer lebhaften Diskussion Veranlassung. An derselben
beteiligen sich haupts"achlich Prof. Zermelo, Prof. Stodola, Prof. P. Weiss. Prof. Predig
und Prof. A. Weber.}'' Dazu folgende Anmerkung, die f"ur unser Thema wesentlich ist. Ernst
Zermelo ist der ber"uhmte Mathematiker, nach dem das Auswahlaxiom in der Mengenlehre
benannt wird. Zu dieser Zeit war er Mathematik Professor an der Universit"at. In jungen
Jahren war er Planck's Assistent und hat 1896 einen kurzen, modern anmutenden Beweis des
Poincar\'{e}'schen Wiederkehrsatzes gegeben. Auf dessen Basis griff er mit jugentlicher
Unversch"amtheit Boltzmann an, und wurde dabei von Planck unterst"utzt. Zu dieser Zeit war
Planck ein scharfer Gegner von Boltzmann's statistischer Auffassung der Entropie und damit
des 2. Hauptsatzes der Thermodynamik. Er glaubte nicht an die ``endlichen'' Atome der
kinetischen Gastheorie. Dazu schrieb er am 23. M"arz 1897 an Leo Graetz:
\begin{quote}
``\textit{In dem Hauptpunkt der Frage stehe ich auf Zermelo's Seite, indem ich der Ansicht
bin, dass es prinzipiell ganz aussichtslos ist, die Geschwindigkeit irreversibler Prozesse,
z.B. der Reibung oder W"armeleitung in Gasen, auf wirklich strengem Wege aus der
gegenw"artigen Gastheorie abzuleiten. Denn da Boltzmann selber zugibt, dass sogar die
\emph{Richtung}, in der die Reibung und W"armeleitung wirkt, nur aus
Wahrscheinlichkeitsbetrachtungen zu folgern ist, so w"are v"ollig unverst"andlich, woher es
denn kommt, dass unter allen Umst"anden auch die \emph{Gr"osse} dieser Wirkungen einen ganz
bestimmten Betrag darstellt.'' }
\end{quote}
Darin kommt ein totales Unverst"andnis von Boltzmann's Ideen zum Ausdruck. Planck schloss
sich bekanntlich erst im Dezember 1900 Boltzmann's Auffassungen an, als ihm schliesslich
nichts anderes "ubrigblieb, seine Interpolationsformel f"ur die Energiedichte der Schwarzen
Strahlung theoretisch zu begr"unden.

Bevor ich zum systematischen Teil meines Vortrages komme, m"ochte ich zur Einstimmung noch
ein paar S"atze aus einem Tonband von 1961 zitieren, in welchem Otto Stern meinem verehrten
Lehrer und sp"ateren Kollegen Res Jost "uber Einstein's Prager und zweiten Z"urcher
Aufenthalt erz"ahlt hat\footnote{Otto Stern kam mit eigenen Mitteln nach Prag, um sich
Einstein anzuschliessen. Als Einstein die Professur an der ETH annahm, begleitete er ihn im
Sommer 1912 nach Z"urich. Dort habilitierte sich Stern im Jahre 1913 in physikalischer
Chemie. Mit einem eingehenden Gutachten hat Einstein diese Habilitation unterst"utzt.}.
Diese geben ein sehr lebendiges Bild von Einstein. (Das vollst"andige Tonband befindet sich
im Archiv der ETH.)

\begin{quote}
``\textit{Einstein war in Prag v"ollig vereinsamt, trotzdem es vier Hochschulen gab: eine
deutsche Universit"at, eine tschechische Universit"at, eine deutsche technische Hochschule,
eine tschechische technische Hochschule. An keiner war ein Mensch, mit dem Einstein "uber
die Sachen sprechen konnte, die ihn wirklich interessierten. Er tat es also nolens volens
mit mir (...). Der einzige wirklich intelligente Mann dort, das war ein Mathematiker namens
Pick. Der war mal Assistent bei Mach gewesen und hatte von da her die "Uberzeugung, dass
Molek"ule einfach Aberglauben w"aren. Also, wenn Einstein und ich "uber die Molek"ule
sprachen, da lachte er uns einfach aus (...). In Z"urich war's nat"urlich sehr sch"on (...)
und besonders deswegen interessant, weil Laue an der Universit"at war. Ausserdem waren
Ehrenfest und Tatjana (...) mindestens ein Vierteljahr, vielleicht auch etwas l"anger zu
Besuch (...). Das gab nat"urlich immer herrliche Diskussionen im Kolloquium (...). Wir
waren auch ein paar j"ungere Leute, die ganz eifrig waren. Ehrenfest nannte uns immer den
`Dreistern'. Das waren der Herzfeld und der Kern (und ich). (Kern hatte den Doktor bei
Debye gemacht.) Debye war ja der Vorg"anger von Laue an der Universit"at (...). Nur der
Weiss (...), Pierre Weiss war damals Experimentalphysiker und Institutsdirektor, der kam
nie ins Kolloquium. Er verbot auch das Rauchen, das war furchtbar (...). Dem Einstein
konnte man das aber nicht verbieten. Infolgedessen, wenn es eben zu schlimm war, dann bin
ich einfach ins Einstein'sche Zimmer gegangen (...) und konnte mich mit ihm unterhalten
(...). Das gab dann immer lebhafte Diskussionen (...) "uber damals v"ollig ungel"oste
R"atsel der Quantentheorie. Das einzige, was man "uber Quantentheorie wirklich wusste, war
die Planck'sche Formel, Schluss (...). Ich bin auch ins Kolleg zu Einstein gegangen (...),
das war (...)auch sehr sch"on, aber nicht f"ur Anf"anger. Einstein hat sich ja nie richtig
vorbereitet auf die Vorlesung, aber er war eben doch Einstein (...), wenn er da so
herumgemorkst hat, war es doch sehr interessant (...), immer sehr raffiniert gemacht und
sehr physikalisch vor allen Dingen (...). Im Winter kamen Planck und Nernst, um Einstein
die Stellung in Berlin anzubieten (...). Einstein sagte mir: `Wissen Sie, die beiden, die
kommen mir vor (...) wie Leute, die seltene Briefmarken erwerben wollen.}''
\end{quote}

Zum Gesamteindruck des Gespr"achs mit Otto Stern meinte Res Jost, dieses habe den Anschein
erweckt, als ob Einstein sich ausschliesslich mit dem Quantenproblem befasst h"atte. Dabei
deckt Stern's Bericht die Zeit ab, in der sich Einstein mit voller Kraft auf den Weg zur
Allgemeinen Relativit"atstheorie gemacht hat. Die Quantenr"atsel liessen ihn eben nie los.

\section{Statistische Mechanik und Lichtquanten}

Ich beginne den systematischen Teil meines Vortrags mit Einstein's erster Arbeit von 1905.
Dar"uber gibt es leider bis heute -- auch von sehr bekannten Leuten -- immer wieder
verkehrte "Ausserungen.

Nach den "uberw"altigenden Erfolgen der Wellentheorie des Lichtes war Einstein's
Lichtquantenhypothese sehr revolution"ar, wie er selber in einem aufschluss\-reichen Brief
an Conrad Habicht schreibt:
\begin{quote}
``(...) \textit{Ich verspreche Ihnen vier Arbeiten daf"ur, von denen ich die erste in
B"alde schicken k"onnte, da ich die Freiexemplare baldigst erhalten werde. Sie handelt
"uber Strahlung und die energetischen Eigenschaften des Lichtes und ist sehr revolution"ar,
wie Sie sehen werden} (...).''
\end{quote}

Auch unter Physikern ist leider die falsche Meinung weit verbreitet, dass sich Einstein in
der Arbeit ``\textit{"Uber einen die Erzeugung und Umwandlung des Lichtes betreffenden
heuristischen Standpunkt}'' in erster Linie mit einer Erkl"arung des photoelektrischen
Effektes befasst habe, wof"ur er 1922 den Nobelpreis erhielt.

In Wirklichkeit war dies lediglich eine der Anwendungen einer viel tieferen Analyse,
aufgrund derer  Einstein zur "Uberzeugung gelangte, dass die Strahlung -- neben ihren
etablierten Welleneigenschaften -- auch eine korpuskulare Struktur besitzt. Diese Arbeit
geh"ort zu Einstein's allergr"ossten Taten. Der eigentliche Kern der Abhandlung besteht in
einer statistisch-mechanischen Analyse der thermodynamischen Gleichgewichtsstrahlung
(Schwarze Strahlung). Dabei konzentriert sich Einstein auf das Wiensche Gebiet hoher
Frequenzen und tiefer Temperaturen, denn er erkennt, dass die klassische Physik mit der
empirischen spektralen Verteilung in diesem Bereich in krassem Widerspruch steht.

Einsteins Lichtquanten wurden f"ur lange Zeit nicht akzeptiert. Auch Planck hielt sie f"ur
allzu k"uhn und radikal. Die eigentliche Anerkennung fand das Lichtquant erst 1922, als man
es geradezu handgreiflich beim Comptoneffekt beobachten konnte. Damit liess sich
schliesslich auch Niels Bohr, der letzte engagierte Gegner der Quantennatur des Lichtes
"uberzeugen.

Wir wollen nun den Gedankengang der Arbeit vom 17. M"arz, deren ``historische Bedeutung und
Originalit"at gar nicht "ubersch"atzt werden k"onnen'' (Res Jost), n"aher betrachten.

In einem ersten Abschnitt betont Einstein, dass die klassische Physik ein unsinniges Gesetz
f"ur die Strahlungsdichte $\rho(T,\nu)$ impliziert, dieses jedoch f"ur grosse Wellenl"angen
und Strahlungsdichten gelten muss. Unter Benutzung des "Aquipartitionstheorems f"ur
Resonatoren in Wechselwirkung mit dem klassischen Strahlungsfeld, findet er unabh"angig das
Rayleigh-Jeans-Gesetz $\rho(\nu,T)=(8\pi\nu^2/c^3)kT$ und betont, dass die zugeh"orige
totale Energiedichte divergiert. Dann stellt er im zweiten Abschnitt fest, dass die
empirisch erfolgreiche Planck'sche Formel tats"achlich im erw"ahnten Grenzbereich damit
"ubereinstimmt, wenn die Avogadro- (Loschmidt-) Zahl den folgenden Wert hat
\begin{equation}
N_A=6.17\times 10^{23},
\end{equation}
welcher bereits von Planck angegeben wurde. Zu seinem Korrespondenzargument betont Einstein
mit Recht, ``dass die von Herrn Planck gegebene Bestimmung der Elementarquanta von der von
ihm aufgestellten Theorie der schwarzen Strahlung unabh"angig ist''. Tats"achlich wusste
Einstein, im Unterschied zu Planck, bei seiner Bestimmung der Loschmidt'schen Zahl, aus
ersten Prinzipien was er tat. Als Fazit betont er nochmals:
\begin{quote}
``\textit{Je gr"osser die Energiedichte und die Wellenl"ange einer Strahlung ist, als um so
brauchbarer erweisen sich die von uns benutzten theoretischen Grundlagen; f"ur kleine
Wellenl"angen und kleine Strahlungsdichten versagen dieselben vollst"andig.}''
\end{quote}

Nun geht Einstein daran zu analysieren, was die Strahlungsformel
\begin{equation}
\rho(T,\nu)=\frac{8\pi\nu^2}{c^3}h\nu e^{-h\nu/kT}
\end{equation}
im Wien'schen Teil "uber das Licht aussagt. Es sei $E_V(T,\nu)$ die Energie der Strahlung
im Volumen $V$ im kleinen Frequenzintervall $[\nu,\nu+\Delta\nu]$,
\begin{equation}
E_V=\rho(T,\nu)V\Delta\nu
\end{equation}
und $S_V=\sigma(T,\nu)V\Delta\nu$ die entsprechende Entropie. Nach der Thermodynamik gilt
\[ \frac{\partial\sigma}{\partial\rho}=\frac{1}{T}. \]
Hier setzen wir f"ur $1/T$ den Ausdruck ein, der sich aus der Wien'schen Strahlungsformel
(2) ergibt und finden
\begin{equation}
\frac{\partial\sigma}{\partial\rho}=-\frac{k}{h\nu}\ln\left[\frac{\rho}{8\pi
h\nu^3/c^3}\right].
\end{equation}
Nach Integration erhalten wir die Beziehung
\begin{equation}
S_V=-k\frac{E_V}{h\nu}\left\{\ln\left[\frac{E_V}{V\Delta\nu\,8\pi
h\nu^3/c^3}\right]-1\right\}.
\end{equation}

Einstein interessiert sich prim"ar f"ur die Volumenabh"angigkeit dieses Ausdrucks f"ur die
Entropie. F"ur diese erh"alt man f"ur einen festen Wert $E=E_V$ der Energie
\begin{equation}
S_V-S_{V_0}=k\frac{E}{h\nu}\ln\left(\frac{V}{V_0}\right)=k\ln\left(\frac{V}{V_0}\right)^{
E/h\nu}.
\end{equation}

Soweit wurde nur die Thermodynamik benutzt. An dieser Stelle bringt nun Einstein das
``Boltzmann'sche Prinzip'' ins Spiel, welches bereits in seinen fr"uhen statistisch
mechanischen Arbeiten eine zentrale Rolle spielte. Nach Boltzmann ist die Entropie $S$
eines Systems gem"ass
\begin{equation}
S=k\ln W,
\end{equation}
mit der Zahl der M"oglichkeiten $W$ verkn"upft, durch die ein makroskopischer Zustand
verwirklicht werden kann. In einem separaten Abschnitt erinnert er an diesen fundamentalen
Zusammenhang zwischen Entropie und ``statistischer Wahrscheinlichkeit'' (Einstein's
Terminologie) und wendet ihn auf ein ideales Gas von $N$ Teilchen an, die in den Volumina
$V,V_0$ eingesperrt sind. F"ur die relative Wahrscheinlichkeit der beiden Situationen gilt
\begin{equation}
W=\left(\frac{V}{V_0}\right)^N,
\end{equation}
und somit f"ur die Entropien
\begin{equation}
S(V,T)-S(V_0,T)=kN\ln\left(\frac{V}{V_0}\right).
\end{equation}

Auf die Formel (6) angewandt, gibt das Boltzmann'sche Prinzip f"ur die Strahlung
\begin{equation}
W=\left(\frac{V}{V_0}\right)^{ E/h\nu}.
\end{equation}
Hier treten, beim Vergleich mit (9), geradezu handgreiflich die Energiequanten $h\nu$ als
Teilchen (Lichtquanten) auf.

Einstein beschliesst diese Betrachtungen mit den ber"uhmten Worten:
\begin{quote}
``\textit{Monochromatische Strahlung von geringer Dichte (innerhalb des
G"ultigkeitsbereichs der Wien'schen Strahlungsformel) verh"alt sich in w"armetheoretischer
Beziehung so, wie wenn sie aus voneinander unabh"angigen Energiequanten der Gr"osse $h\nu$
best"unde.}''
\end{quote}

Diese Einsicht ist eine Frucht der Statistischen Mechanik und noch keine revolution"are
Aussage.

\subsubsection*{Lichtquantenhypothese}

Einstein's revolution"arer Schritt besteht in einer Aussage "uber die Quanteneigenschaften
der freien elektromagnetischen Strahlung, die f"ur lange Zeit von sonst niemandem
akzeptiert wurde. Er formuliert sein \textit{heuristisches Prinzip} folgendermassen:
\begin{quote}
``\textit{Wenn sich nun monochromatische Strahlung (von hinreichend kleiner Dichte)
bez"uglich der Abh"angigkeit der Entropie vom Volumen wie ein diskontinuierliches Medium
verh"alt, welches aus Energiequanten von der Gr"osse $h\nu$ besteht, so liegt es nahe zu
untersuchen, ob auch die Gesetze der Erzeugung und Verwandlung des Lichtes so beschaffen
sind, wie wenn das Licht aus derartigen Energiequanten best"unde.}''
\end{quote}

In zwei Schlussabschnitten wendet Einstein seine Hypothese auf die Stoke'sche Regel bei der
Photolumineszenz und den lichtelektrischen Effekt an. F"ur letzteren sagt er die bekannte
lineare Beziehung
\begin{equation}
E_{max}=h\nu-P
\end{equation}
zwischen der maximalen Energie der Photoelektronen ($E_{max}$)und der Frequenz $\nu$ der
einfallenden Strahlung voraus, welche erst zehn Jahre sp"ater durch Millikan best"atigt
wurde. (In dieser ber"uhmten Gleichung bezeichnet $P$ die Austrittsarbeit.) Als Einstein
seine Arbeit schrieb war "uber den lichtelektrischen Effekt nur Qualitatives bekannt.
Deshalb war die lineare Beziehung (11) zwischen $E_{max}$ und $\nu$ -- mit der Steigung $h$
-- eine \emph{Voraussage}.

\subsection*{Energie- und Impulsschwankungen}

Ein paar Jahre sp"ater hat Einstein die besprochene statistisch mechanische Analyse noch
mit der Berechnung der Energie- und Impulsschwankungen der Schwarzen Strahlung vertieft. An
Hand der resultierenden ber"uhmten Schwankungsformeln, mit zwei unterschiedlichen
Beitr"agen f"ur das klassische, bzw. Wien'schen Regime, machte er nachdr"ucklich auf eine
merkw"urdige Doppelnatur der Strahlung aufmerksam. Ich erinnere kurz an eine der
wegweisenden Schwankungsformeln, die Einstein in seiner Arbeit ``Zum gegenw"artigen Stand
des Strahlungsproblems'' aus dem Jahre 1909 herleitet. Dabei benutzt er seine allgemeine
Energie-Schwankungsformel im kanonischen Ensemble, welche er in seiner dritten Arbeit "uber
Statistische Mechanik erhalten hatte. Diese ergibt f"ur die Varianz von $E_V$ in (3)
\begin{equation}
\left\langle(E_V-\langle E_V\rangle)^2\right\rangle= kT^2\frac{\partial\langle
E_V\rangle}{\partial T}=kT^2V\Delta\nu\frac{\partial\rho}{\partial T}.
\end{equation}
F"ur die Planck-Verteilung liefert dies
\begin{equation}
\left\langle(E_V-\langle
E_V\rangle)^2\right\rangle=\left(h\nu\rho+\frac{c^3}{8\pi\nu^2}\rho^2\right)V\Delta\nu.
\end{equation}
Den zweiten Term rechts w"urde die Jeans-Formel ergeben und ist deshalb erwartungsgem"ass
ein Ausdruck f"ur die Wellennatur des Lichtes. Einstein zeigt durch eine
Dimensionsbetrachtung, welche sp"ater durch H.A. Lorentz in quantitativer Weise erg"anzt
wurde, dass dieser Schwankungsanteil tats"achlich durch Interferenz der Lichtwellen mit
ann"ahernd gleichen Wellenvektoren (Schwebungen) zustande kommt. Hingegen ist der erste
Term ein Ausdruck f"ur die Teilchennatur der Strahlung.

\subsection*{Reaktionen}

Da die Beitr"age von Planck und Einstein nach wie vor oft durcheinander gebracht werden,
will ich diesen Teil des Vortrags mit Aussagen von Planck beschliessen, die deutlich
zeigen, dass er auch noch viele Jahre sp"ater die Lichtquanten ablehnte. In einem Brief von
Planck vom 6. Juli 1907 steht der Satz: ``Ich suche die Bedeutung des elementaren
Wirkungsquantums nicht im Vakuum, sondern an der Stelle der Absorption u. Emission, und
nehme an, dass die Vorg"ange im Vakuum durch die Maxwell'schen Gleichungen \emph{genau}
dargestellt werden.'' Als Planck, Nernst, Rubens und Wartburg im Jahre 1913 Einstein f"ur
die Mitgliedschaft in der Preussischen Akademie vorschlugen, beschlossen sie ihre
Stellungsnahme -- nach h"ochstem Lob -- wie folgt:

\begin{quote}
``\textit{Zusammenfassend kann man sagen, dass es unter den grossen Problemen, an denen die
moderne Physik so reich ist, kaum eines gibt, zu dem nicht Einstein in bemerkenswerter
Weise Stellung genommen h"atte. Dass er in seinen Spekulationen gelegentlich auch einmal
"uber das Ziel hinausgeschossen haben mag, wie z.B. in seiner Hypothese der Lichtquanten,
wird man ihm nicht allzuschwer anrechnen d"urfen; denn ohne einmal ein Risiko zu wagen,
l"asst sich auch in der exaktesten Naturwissenschaft keinerlei wirkliche Neuerung
einf"uhren.}''
\end{quote}

Als weiteres Beispiel f"ur die allgemeine Ablehnung von Einstein's Lichtquantenhypothese
sei noch Millikan zitiert. Nachdem er mit seinen ber"uhmten Experimenten Einstein's Formel
(11) sehr genau best"atigt hatte -- das Planck'sche $h$ wurde dabei auf ein halbes Prozent
genau bestimmt -- schrieb er
\begin{quote}
``...\textit{trotz des offensichtlich vollst"andigen Erfolgs der Einstein-Gleichung
erscheint die physikalische Theorie dessen, wof"ur die Gleichung der symbolische Ausdruck
sein soll, so unhaltbar, dass nach meiner Meinung Einstein selbst nicht l"anger daran
glaubt.}''
\end{quote}

\section{Einstein's Dissertation}

Die zweite der f"unf Arbeiten von 1905 tr"agt den Titel ``\textit{Eine neue Bestimmung der
Molek"uldimensionen}''. Sie wurde am 30. April fertiggestellt und am 20. Juli dem Dekan der
Philosophischen Fakult"at II der Universit"at Z"urich vorgelegt. Einstein's Dissertation
ist weniger bekannt als die anderen vier Arbeiten, zu Unrecht wie mir scheint. Die
ber"uhmtere Untersuchung "uber Brownsche Bewegung schloss direkt daran an und traf nur elf
Tage sp"ater, am 11. Mai bei den \textit{Annalen der Physik} ein.

Doktorpromotionen wurden an der ETH, dem damaligen `Polytechnikum', erst ab Herbst 1909
m"oglich. Dies ist einer der Gr"unde, weshalb Einstein seine Promotionsgesuch an der
Universit"at einreichte. Dort war Alfred Kleiner der einzige Lehrstuhlinhaber in Physik.
Seine Forschung war der Entwicklung von Messinstrumenten gewidmet, aber er hatte auch ein
gewisses Interesse f"ur  Grundlagenphysik. Dies geht z.B. aus  folgenden Bemerkungen von
Einstein aus einem Brief an Mileva vom 19. Dezember 1901 hervorgeht: ``Heute war ich den
ganzen Nachmittag beim Kleiner in Z"urich und hab ihm meine Ideen zur Elektrodynamik
bewegter K"orper erkl"art \& auch sonst "uber alle m"oglichen physikalischen Fragen mit ihm
gesprochen. Er ist doch nicht ganz so dumm wie ich gemeint habe, und vor allem, er ist ein
guter Kerl.'' Einstein zeigte Kleiner auch seine erste Dissertation vom November 1901.
Diese hat nicht "uberlebt und es ist nicht klar was sie enthielt. Einstein zog die Arbeit
im Februar 1902 zur"uck. Eine zeitlang gab er den Gedanken an eine Promotion auf, ``da mir
das doch wenig hilft und die ganze Kom"odie mir langweilig ist''.

Kleiner war nat"urlich einer der Gutachter von Einstein's Promotionsarbeit. Sein Urteil
fiel sehr positiv aus: ``die "Uberlegungen und Rechnungen geh"oren zu den schwierigsten der
Hydrodynamik.'' Der andere Gutachter, Heinrich Burkhard, Professor f"ur Mathematik an der
Universit"at, meinte: ``die Art der Behandlung zeugt von \emph{gr"undlicher Beherrschung
der in Frage kommenden mathematischen Methoden.}''


Ich bespreche nun zuerst den wesentlichen Inhalt der Dissertation und werde anschliessend
vorallem auf den hydrodynamischen Teil etwas genauer eingehen.

Einstein studiert in seiner -- Marcel Grossmann gewidmeten -- Dissertation suspendierte
Teilchen, z.B. grosse (Zucker-) Molek"ule, in einer Fl"ussigkeit. Unter der Annahme, dass
die L"osung verd"unnt ist, berechnet er in einem ersten Teil die relative "Anderung der
Viskosit"at (relativ zum L"osungsmittel) und zeigt, dass diese durch das relative Volumen
der suspendierten Teilchen bestimmt ist. (Diese Rechnung ist, wie wir sehen werden, gar
nicht einfach und beansprucht die meisten Druckseiten. Einstein verwendet dabei Kirchhoffs
Vorlesungen "uber mathematische Physik, die er seit seinen Studententagen kannte.) Sp"ater
wendet er seine Formel auf w"assrige Zuckerl"osungen an; die gel"osten Teilchen sind dann
die Zuckermolek"ule. Damit erh"alt er eine Beziehung zwischen der Avogadro-Zahl $N_A$ und
der Ausdehnung $a$ der Zuckermolek"ule. Danach sucht er nach einer zweiten Beziehung, die
kurz darauf auch in der Arbeit "uber Brownsche Bewegung wichtig wurde. Mit einer
bestechenden "Uberlegung gelingt es ihm, die Diffusionskonstante ebenfalls durch $N_A$ und
$a$ auszudr"ucken. Aus dem Vergleich dieser Ergebnisse mit Daten "uber w"assrige
Zuckerl"osungen fand er auf diesem Weg einen guten Wert f"ur die Avogadro'sche Zahl sowie
f"ur die Ausdehnung der Zuckermolek"ule. Dies war ein erster wichtiger Schritt in Einstein'
Bem"uhungen, weitere Evidenz f"ur die Atomhypothese zu liefern. (Tats"achlich enthielt die
Dissertation noch einen trivialen Rechenfehler, der erst einige Jahre sp"ater behoben
wurde. Damit sowie besseren experimentellen Daten ergab sich eine bemerkenswerte
"Ubereinstimmung mit den Resultaten von anderen Methoden, insbesondere mit dem Ergebnis von
Jean Perrin, das auf der Basis von Einsteins Theorie der Brownschen Bewegung gewonnen
wurde. Mehr dazu sp"ater.)

Hier ist anzumerken, dass die physikalisch Realit"at der Atome  am Ende des 19.
Jahrhunderts noch nicht universell akzeptiert wurde. Leidenschaftliche und einflussreiche
Gegner waren u.a. Wilhelm Ostwald und Georg Helm, und nat"urlich Ernst Mach. Letzterer
betrachtete den Atomismus lediglich als didaktisches oder heuristisches Werkzeug. Es wird
berichtet, dass Mach jeden der von Atomen sprach mit der Standardfrage nervte: ``Ham's eins
g'sehn?''.

Wie andere bedeutende Arbeiten, hatte auch Einstein's Dissertation eine gewisse
Inkubationszeit hinter sich. So nehmen die folgenden Bemerkungen in einem  Brief an Besso
vom 17. M"arz 1903 schon wesentliche Elemente vorweg, vor allem im zweiten Teil:
\begin{quote}
`` \textit{Hast Du die absolute Gr"osse der Ionen schon ausgerechnet unter der
Vorausetzung, dass dieselben Kugeln und so gross sind, dass die Gleichungen der
Hydrodynamik reibender Fl"ussigkeiten anwendbar sind. Bei unserer Kenntnis der absoluten
Gr"osse des Elektrons w"are dies ja eine einfache Sache. Ich h"atte es selbst getan, aber
es fehlt mir Literatur und Zeit; auch die Diffusion k"onntest Du heranziehen, um "uber die
neutralen Salzmolek"ule in L"osung Aufschluss zu erhalten.}''
\end{quote}

Bevor ich auf die Einzelheiten der Dissertation eingehe, m"ochte ich noch auf ein weiteres
Motiv hinweisen, das f"ur Einstein wichtig war. Ein paar Jahre sp"ater schrieb er dazu an
Perrin:

\begin{quote}
``\textit{Eine pr"azise Bestimmung der Gr"osse der Molek"ule scheint mir deshalb von
h"ochster Wichtigkeit, weil durch eine solche die Strahlungsformel von Planck sch"arfer
gepr"uft werden kann als durch Strahlungsmessungen}.''
\end{quote}

Zu einem bedeutenden Wissenschafter geh"ort -- neben Originalit"at und Intuition -- auch
eine gute Technik. Die nachstehenden Ausf"uhrungen sollten zeigen, dass dies bei Einstein
bereits in fr"uhen Jahren der Fall war.

\subsection{Hydrodynamische Grundlagen}

Im folgenden benutze ich heutige Vektor- (und Tensor-) Schreibweisen.

F"ur station"are Str"omungen einer inkompressiblen homogenen Fl"ussigkeit lautet die
Navier-Stokes-Gleichung in Standardbezeichnungen
\[ (\mathbf{v}\cdot\nabla)\mathbf{v}=-\frac{1}{\rho}\nabla
p+\frac{\eta}{\rho}\triangle\mathbf{v}.\] F"ur kleine Reynold-Zahlen kann man die linke
Seite vernachl"assigen. In dieser Situation lauten demnach die Grundgleichungen
\begin{equation}
\nabla p=-\eta\triangle\mathbf{v}, ~~~\nabla\cdot\mathbf{v}=0.
\end{equation}
Als Folge davon ist der Druck harmonisch: $\triangle p=0$. Dasselbe gilt auch f"ur die
Vortizit"at $\mathrm{rot}~\mathbf{v}$. Wir erinnern auch an den Ausdruck f"ur den
Spannungstensor
\begin{equation}
\sigma_{ij}=-p\delta_{ij}+\eta(\partial_iv_j+\partial_jv_i),
\end{equation}
der nach (14) divergenzfrei ist: $\partial_j\sigma_{ij}=0$. Sp"ater ben"otigen wir auch die
folgende Formel f"ur die Arbeitsleistung $W$, welche die Spannungen auf die Oberfl"ache
$\partial\Omega$ eines Gebietes $\Omega$ aus"uben
\begin{equation}
W=\int_{\partial\Omega}v_i\sigma_{ij}n_j~dA.
\end{equation}
Darin ist $\mathbf{n}$ der Einheitsvektor in Richtung der "ausseren Normalen.

\subsection{Einstein's Strategie}

Mit Einstein  betrachten wir nun eine inkompressible Fl"ussigkeit der Z"ahigkeit $\eta_0$,
in der eine Vielzahl identischer kleiner fester Teilchen suspendiert sind. Die Suspension
kann auf zwei unterschiedliche Arten beschrieben werden: 1. Als homogenes Medium mit einer
effektiven Viskosit"at $\eta$ auf Skalen, die im Vergleich zu den Abmessungen der Teilchen
gross sind. 2. Durch die Str"omung der Fl"ussigkeit (L"osungsmittel), welche von den
suspendierten Teilchen modifiziert wird.

F"ur beide Beschreibungen berechnet nun Einstein gem"ass (16) die Leistung f"ur ein grosses
Gebiet $\Omega$ und erh"alt durch Gleichsetzen die Formel
\begin{equation}
\eta=\eta_0\left(1+\frac{5}{2}\varphi\right),
\end{equation}
wo $\varphi$ den Bruchteil des Volumens bezeichnet, den die suspendierten Teilchen
einnehmen. Dieser wird als klein vorausgesetzt (verd"unnte Suspension). (Als Folge eines
Rechenfehlers hatte Einstein urpr"unglich den Faktor 5/2 verloren; mehr zu dieser
am"usanten Geschichte sp"ater.)

\subsection{Geschwindigkeitsfeld f"ur ein einzelnes suspendiertes Teilchen}

Wir verwenden zuerst die zweite Beschreibung und m"ussen dann als vorbereitende Aufgabe den
Einfluss berechnen, den ein einzelnes in die Fl"ussigkeit eingetauchtes K"ugelchen auf eine
Str"omung hat, deren Geschwindigkeitsgradient beispielsweise konstant ist. Mathematisch
bedeutet dies die L"osung eines Randwertproblems f"ur das elliptische System (14).

Das Geschwindigkeitsfeld der ungest"orten Str"omung sei also
\begin{equation}
v_i^{(0)}=e_{ij}x_j,
\end{equation}
wobei $e_{ij}$ ein konstanter, symmetrischer und spurloser Tensor ist. Die letzte
Eigenschaft ist eine Folge der Inkompressibilit"at. $e_{ij}$ ist der Deformationstensor (an
Rotationen sind wir nicht interessiert). Den ungest"orten konstanten Druck bezeichnen wir
mit $p^{(0)}$. Der Spannungstensor f"ur die Grundstr"omung ist
\begin{equation}
\sigma_{ij}^{(0)}=-p^{(0)}\delta_{ij}+2\eta_0e_{ij}.
\end{equation}

Die ver"anderte Str"omungsgeschwindigkeit $\mathbf{v}$ zerlegen wir gem"ass
\begin{equation}
\mathbf{v}=\mathbf{v}^{(0)}+\mathbf{v}^{(1)}
\end{equation}
in den ungest"orten Anteil plus einer St"orung $\mathbf{v}^{(1)}$. Die Randbedingungen
sind: $\mathbf{v}=0$ auf der Kugel mit Radius $a$ und $\mathbf{v}=\mathbf{v}^{(0)}$ im
Unendlichen. Analoge Zerlegungen benutzen wir f"ur den Druck und die Spannungen:
\begin{equation}
p=p^{(0)}+p^{(1)}, ~~~\sigma_{ij}=\sigma_{ij}^{(0)}+\sigma_{ij}^{(1)},
\end{equation}
wobei
\begin{equation}
\sigma_{ij}^{(1)}=-p^{(1)}\delta_{ij}+\eta_0(\partial_iv_j^{(1)}+\partial_jv_i^{(1)}).
\end{equation}
F"ur $W$ haben wir dann die Zerlegung ($|\Omega|$= Volumen von $\Omega$)
\begin{equation}
W=2\eta_0e_{ij}e_{ij}|\Omega|+e_{ik}\int_{\partial\Omega}\sigma_{ij}^{(1)}x_kn_j~dA
+\int_{\partial\Omega}v_i^{(1)}\sigma_{ij}^{(0)}n_j ~dA.
\end{equation}

Die Kugel legen wir in den Koordinatenursprung. Einstein bestimmt die St"orungen
$v_i^{(1)}$ und $p^{(1)}$ mit Hilfe einer Methode die in Kirchhoff's ``Vorlesungen "uber
Mechanik''\footnote{G. Kirchhoff, \textit{Vorlesungen \"{u}ber mathematische Physik}, Vol.
1, \textit{Mechanik}, Teubner (1897).} beschrieben wird. Diese involviert die beiden
folgenden Schritte: a) Bestimme eine Funktion $V$, welche die Gleichung
\begin{equation}
\triangle V=\frac{1}{\eta_0}p^{(1)}
\end{equation}
erf"ullt, und setze
\begin{equation}
v_i^{(1)}=\partial_iV+v_i',
\end{equation}
wobei die $v_i'$ die folgenden Gleichungen erf"ullen soll
\begin{equation}
\triangle v_i'=0, ~~~\partial_iv_i'=-\frac{1}{\eta_0}p^{(1)}.
\end{equation}
\textit{Anmerkung} zu a). Falls dies gelungen ist, sind die Grundgleichungen f"ur
$v_i^{(1)}$ und $p^{(1)}$ erf"ullt:
\[ \eta_0\triangle v_i^{(1)}=\eta_0\partial_i\triangle V=\partial_ip^{(1)}, ~~
\partial_iv_i^{(1)}=\triangle V+\partial_iv_i'=0.\]
b) F"ur $p^{(1)}$ mache man den abfallenden harmonischen Ansatz
\begin{equation}
\frac{p^{(1)}}{\eta_0}=Ae_{ij}\partial_i\partial_j\left(\frac{1}{r}\right)
\end{equation}
mit einer Konstanten $A$, und f"ur $v_i'$ versuche man den harmonischen Ausdruck
\begin{equation}
v_i'=-\tilde{A}e_{ik}\partial_k\left(\frac{1}{r}\right)
+B\partial_ie_{jk}\partial_j\partial_k\left(\frac{1}{r}\right).
\end{equation}
Letzterer erf"ullt beide Gleichungen (26) f"ur $\tilde{A}=A$, denn dann haben wir
\[\partial_iv_i'=-Ae_{ik}\partial_i\partial_k\left(\frac{1}{r}\right)=-\frac{p^{(1)}}{\eta_0}.\]
Die Gleichung (24) ist wegen $\triangle r=2/r$ erf"ullt f"ur
\begin{equation}
V=\frac{1}{2}Ae_{ij}\partial_i\partial_j r.
\end{equation}
Ausdifferenzieren von (28) und (29) liefert

\begin{equation}
v_i^{(1)}=\frac{3}{2}Ae_{jk}\frac{x_ix_jx_k}{r^5}+
B\left(6e_{ik}\frac{x_k}{r^5}-15e_{jk}\frac{x_ix_jx_k}{r^7}\right).
\end{equation}
Die Randbedingung $v_i^{(1)}=-e_{ij}x_j$ f"ur $r=a$ erfordert
\begin{equation}
A=-\frac{5}{3}a^3, ~~~B=-\frac{a^5}{6}.
\end{equation}
Somit erhalten wir f"ur die St"orung des Geschwindigkeitsfeldes ($n_i:=x_i/r$)
\begin{equation}
v_i^{(1)}=-\frac{5}{2}a^3e_{jk}\frac{1}{r^2}n_in_jn_k-
\frac{a^5}{6}\left(6e_{ik}\frac{x_k}{r^5}-15e_{jk}\frac{x_ix_jx_k}{r^7}\right).
\end{equation}
Nach (25) und (28) f"ur $\tilde{A}=A$ k"onnen wir $v_i^{(1)}$ auch folgendermassen
darstellen
\begin{equation}
v_i^{(1)}=-\frac{5}{6}a^3e_{jk}\partial_i\partial_j\partial_k (r)
+\frac{5}{3}a^3e_{ik}\partial_k\left(\frac{1}{r}\right)
-\frac{1}{6}a^5\partial_ie_{jk}\partial_j\partial_k\left(\frac{1}{r}\right).
\end{equation}
F"ur den Druck gibt (27)
\begin{equation}
p=p^{(0)}-5\eta_0a^3e_{ij}\frac{n_in_j}{r^3}.
\end{equation}

Einstein behauptet, dass die Eindeutigkeit der L"osung seines Randwertproblems gezeigt
werden k"onne, aber er macht dazu lediglich ein paar Andeutungen eines m"oglichen Beweises.
Offensichtlich war ihm nicht bekannt, dass ein eleganter Eindeutigkeitsbeweis f"ur solche
Probleme bereits 1868 von Helmholtz\footnote{H. Helmholtz, ``Theorie der station\"{a}ren
Str\"{o}me in reibenden Fl\"{u}ssigkeiten'', Wiss. Abh., Bd. I, S. 223.} gegeben wurde.
Dieser verl"auft folgendermassen: F"ur zwei L"osungen der Grundgleichungen, mit gegebenen
Werten der Geschwindigkeitsfelder auf den R"andern, betrachte man die nicht-negative
Gr"osse $(\theta'_{ij}-\theta_{ij})(\theta'_{ij}-\theta_{ij})$, wobei
$\theta'_{ij},\theta_{ij}$ die Deformationstensoren der beiden Geschwindigkeitsfelder
bezeichnen. Es ist leicht zu zeigen, dass das Integral dieser Funktion "uber das Gebiet
ausserhalb der K"orper verschwindet. (Dies ergibt sich nach zweimaliger Anwendung des
Gauss'schen Integralsatzes, unter Benutzung der Grundgleichungen inklusive der
Randbedingungen.) Deshalb gilt $\theta'_{ij}=\theta_{ij}$; mit anderen Worten, muss der
Deformationstensor f"ur die Differenz $v'_i-v_i$ der beiden Geschwindigkeitsfelder
verschwinden. Deshalb entspricht diese Differenz eine Kombination von einer starren
Translation und einer starren Rotation. Aufgrund der Randbedingungen m"ussen damit die
beiden Geschwindigkeitsfelder "ubereinstimmen. Bis auf eine irrelevante additive Konstante,
trifft dies dann auch f"ur die Dr"ucke der beiden L"osungen zu.

\subsection{Zwei Ausdr"ucke f"ur die Leistung $W$}

In (23) w"ahlen wir f"ur $\Omega$ nun eine grosse Kugel $K_R$ mit dem Radius $R$. In
f"uhrender Ordnung tr"agt nur der erste Term von (32) bei, und eine Routine-Rechnung
liefert den folgenden Ausdruck f"ur $W$ in den sph"arischen Momenten
\[
\overline{n_in_jn_kn_l}:=\frac{1}{4\pi}\int_{S^2}n_in_jn_kn_l
~d\Omega=\frac{1}{15}(\delta_{ij}\delta_{kl}+
\delta_{ik}\delta_{jl}+\delta_{il}\delta_{lk}),\]
\begin{eqnarray*}
W &=& 2\eta_0e_{ij}e_{ij}|\Omega|+20\pi
a^3\eta_0e_{ik}\{3e_{rs}\overline{n_in_kn_rn_s}-e_{is}\overline{n_sn_k}\} \\
&=& 2\eta_0e_{ij}e_{ij}\left\{|\Omega|+\frac{1}{2}\frac{4\pi}{3}a^3\right\}.
\end{eqnarray*}
Dies gilt f"ur eine einzelne Kugel. Solange die Suspension verd"unnt ist, erhalten wir
Einstein's (korrigiertes) Resultat
\begin{equation}
W=2\eta_0e_{ij}e_{ij}|\Omega|\left(1+\frac{1}{2}\varphi\right).
\end{equation}

Nun berechnen wir mit Einstein dieselbe Gr"osse in der ersten Beschreibung  der Suspension.
Dazu schreiben wir das Resultat (33) in der Form
\begin{equation}
v_i=e_{ij}x_j+(e_{ik}\triangle-e_{jk}\partial_i\partial_j)\partial_kf,
\end{equation}
mit
\begin{equation}
f=-\frac{1}{2}Ar-\frac{B}{r}
\end{equation}
($A$ und $B$ haben die fr"uhere Bedeutung (31)). Wenn wir nun die Beitr"age von allen
suspendierten K"ugelchen mit Anzahldichte $n$ im Ball $K_R$ aufsummieren, so erhalten wir
f"ur das Geschwindigkeitsfeld in $K_R$
\begin{equation}
v_i=e_{ij}x_j+(e_{ik}\triangle-e_{jk}\partial_i\partial_j)\partial_kF,
\end{equation}
wo
\begin{equation}
F(|\mathbf{x}|)=n\int_{K_R}f(|\mathbf{x}-\mathbf{x}'|)~d^3x'=\frac{\pi}{3}nA
\left(\frac{1}{10}r^4-r^2R^2\right)-2\pi n B\left(R^2-\frac{1}{3}r^2\right).
\end{equation}
Daraus findet man leicht
\begin{equation}
v_i=e_{ij}x_j(1-\varphi).
\end{equation}
Dieses Resultat erhielt Einstein leicht anders. Aus ihm ergibt sich ein zweiter Ausdruck
f"ur $W$:
\begin{equation}
W=2\eta e_{ij}e_{ij}|\Omega|(1-2\varphi).
\end{equation}
Durch Vergleich von (35) und (41) folgt das vorweggenommene Hauptresultat (17) von
Einstein.

\subsection{Zwei Beziehungen zwischen Avogadro-Zahl \\ und Molek"ulradius}

Sind nun die harten Kugeln Molek"ule, z.B. Zucker, so ist
\begin{equation}
\varphi=\frac{4\pi}{3}a^3\frac{N_A\rho_s}{m_s},
\end{equation}
wo $\rho_s$ die Dichte der L"osung und $m_s$ deren molare Masse sind, die Einstein bekannt
waren. Ferner gab es Messungen von $\eta/\eta_0$ f"ur verd"unnte Zuckerl"osungen. Somit
erhielt er aus (15) eine Beziehung zwischen $N_A$ und $a$.

Unter Benutzung der vorhandenen Daten f"ur w"assrige Zuckerl"osungen stellt Einstein dabei
folgendes fest.\footnote{Ich gebe hier seine sp"ateren Zahlen von 1911.} Ein Gramm von in
Wasser gel"ostem Zucker hat auf den \textit{Reibungskoeffizienten} denselben Einfluss wie
kleine suspendierte Kugeln vom Gesamtvolumen 0.98 $cm^3$. Anderseits verh"alt sich die
\textit{Dichte} von Zuckerl"osung experimentell wie eine Mischung von Wasser und Zucker in
gel"oster Form von spezifischem Volumen 0.61 $cm^3$. (Letzteres ist auch das Volumen von
einem Gramm festen Zuckers.) Den Unterschied der beiden Zahlen interpretiert Einstein als
Anlagerung von Wassermolek"ulen an die Zuckermolek"ule. Der Radius $a$ in (40) ist demnach
ein ``hydrodynamisch wirksamer Molek"ulradius'', der die Vergr"osserung durch die
Hydratation ber"ucksichtigt.

\subsubsection*{Diffusion}

Um die beiden Gr"ossen individuell bestimmen zu k"onnen, suchte Einstein nach einer zweiten
Beziehung und fand dabei seine ber"uhmte Diffusionsformel, welche danach auch in der Arbeit
"uber Brown'sche Bewegung eine Schl"usselrolle spielen sollte. Die Herleitung dieser Formel
ist kurz, aber bestechend (``ungeheuer einfallsreich'', wie Pais sagt). Sie beruht auf
thermischen und mechanischen Gleichgewichtsbetrachtungen.

Auf die suspendierten Teilchen der Anzahldichte $n$ wirke eine konstante "aussere Kraft
$\mathbf{f}$ pro Teilchen. Diese verursacht einen Teilchenstrom der Gr"osse $n\mathbf{v}$
($\mathbf{v}=$ Geschwindigkeit des Teilchenstroms). Im Gleichgewicht wird dieser durch den
Diffusionsstrom $-D\nabla n,~D=$ Diffusionskonstante, kompensiert. Die Geschwindigkeit des
Teilchenstromes ist proportional zu $\mathbf{f}$,
\begin{equation}
\mathbf{v}=b\mathbf{f},~~~b:\mathnormal{Beweglichkeit}.
\end{equation}
Somit erhalten wir eine erste (dynamische) Gleichgewichtsbedingung
\begin{equation}
D\nabla n=nb\mathbf{f}.
\end{equation}

Im thermischen Gleigewicht wird die "aussere Kraft durch den Gradienten des osmotischen
Drucks kompensiert. Nach dem Gesetz von van't Hoff\footnote{Danach ist der osmotische Druck
$p$ der suspendierten Teilchen gleich gross, als wenn sie alleine als ideales Gas vorhanden
w"aren. Deshalb gilt im Gleichgewicht $n\mathbf{f}=\nabla p=kT\nabla n$.} bedeutet dies
\begin{equation}
\mathbf{f}=\frac{kT}{n}\nabla n.
\end{equation}
Setzen wir das in die letzte Beziehung ein, so erhalten wir die einfache Formel
\begin{equation}
D=kTb.
\end{equation}
F"ur $b$ verwendet Einstein die Stokes'sche Formel
\begin{equation}
b=\frac{1}{6\pi\eta_0a}
\end{equation}
und erh"alt so seine wichtige Formel
\begin{equation}
D=\frac{kT}{6\pi\eta_0a},~~k=\frac{R}{N_A}
\end{equation}
($R=$ Gaskonstante).

Diese Formel wurde fast gleichzeitig in Australien von William Sutherland gefunden.
"Ahnliche Gleichgewichtsbetrachtungen zwischen systematischen und fluktuierenden Kr"aften
hat Einstein noch wiederholt angestellt. (Ein wichtiges Beispiel wird in Abschnitt 6
besprochen.)

\subsection{Stillschweigen, ein Rechenfehler,\\sp"ate Beachtung}

Im Jahre 1909 ergaben Perrins genaue Messungen der Brown'schen Bewegung einen neuen Wert
f"ur die Avogadro-Zahl, der jedoch betr"achtlich von demjenigen in Einsteins Dissertation
abwich. Einstein machte daraufhin Perrin auf seine hydrodynamische Methode aufmerksam und
regte an, diese auf die von Perrin benutzten Suspensionen anzuwenden. Die ervorderlichen
Viskosit"atsmessungen wurden sogleich von Jaques Bancelin, einem Sch"uler von Perrin,
durchgef"uhrt. Bancelin best"atigte, dass die Viskosit"at in einer Weise zunimmt die
unabh"angig von der Ausdehnung der suspendierten Teilchen ist und nur von deren
Gesmtvolumen abh"angt. Er erhielt aber ein steileres Anwachsen mit $\varphi$ als von
Einstein vorausgesagt.

Als Folge dieser Entwicklung schrieb Einstein am 27. Dezember 1910  aus Z"urich seinem
fr"uheren Studenten und Mitarbeiter Ludwig Hopf folgendes:

\begin{quote}
``\textit{Das Neue in der Physik ist folgendes. Perrin hat durch einen jungen Physiker
(Jacques Bancelin) die Viskosit"at von Mastixemulsionen experimentell untersuchen lassen.
Er findet\footnote{Es muss hier angemerkt werden, dass an Stelle der Zahl 3.8 (bzw. 3.9 in
einem Brief von Einstein an Perrin) in der Publikation von Jacques Bancelin der Wert 2.9
angegeben wird.} $\eta=\eta_0(1+3.8\varphi)$ (...). Die Untersuchung wurde gemacht, um eine
von mir abgeleitete Formel zu pr"ufen. Diese lautet aber $\eta=\eta_0(1+\varphi)$. Die
Sache ist wichtig, weil man aus der Viskosit"at etwas erfahren kann "uber das Volumen
\emph{gel"oster} Molek"ule.\\ Ich habe nun meine damaligen Rechnungen und "Uberlegungen
gepr"uft und keinen Fehler darin gefunden. Sie w"urden sich sehr um die Sache verdient
machen, wenn Sie meine Untersuchungen seri"os "uberpr"ufen w"urden. Entweder ist ein Fehler
in meiner Arbeit oder das Volumen von Perrin's suspendierter Substanz ist in suspendiertem
Zustand gr"osser als Perrin glaubt. Die Sache ist auch wegen Perrin's Hauptarbeit wichtig,
in der er mit "ahnlichen Suspensionen operierte.}''
\end{quote}

Hopf fand tats"achlich einen Rechenfehler, nach dessen Korrektur sich die Formel (17)
ergab. Dies teilte Einstein sofort Perrin mit, und berichtigte in einer kurzen Mitteilung
in den Annalen der Physik seine Dissertation. (Diese Berichtigung wurde "ubrigens mehr
zitiert als z.B. seine Relativit"atstheorie.) Neue Daten "uber Zuckerl"osungen ergaben nun
den Wert
\begin {equation}
N_A=6.56\times10^{23}
\end{equation}
f"ur die Avogadro-Zahl, in guter "Ubereinstimmung mit den Resultaten von anderen Methoden,
insbesondere mit Perrin's Bestimmungen\footnote{Siehe dazu Perrin's detaillierten Bericht
anl"asslich der Tagung des \emph{Conseil Solvay} von 1911 in: Abhandlungen der Deutschen
Bunsen-Gesellschaft; Dritter Band (S. 125-207),  W. Nernst (Hrsg.).} aus der Brown'schen
Bewegung.

\section{Brown'sche Bewegung}

Die Diffusionsformel (48) spielt auch eine zentrale Rolle in Einstein's Theorie der
Brown'schen Bewegung. Dort leitet er diese nochmals her, wobei gewisse
molekular-theoretische Argumente hinzukommen. Als Hauptpunkt zeigt er in dieser viel
ber"uhmteren Arbeit bekanntlich, dass die h"ochst unregelm"assigen Irrwege der
suspendierten Teilchen -- als Folge unvorstellbar vieler Molek"ulst"osse -- durch einen
station"aren Gauss'schen Prozess beschrieben werden. Dabei wird die Breite der
Wahrscheinlichkeitsverteilung f"ur den Ort durch die Diffusionskonstante bestimmt. Damit
folgt f"ur das 1-dimensionale Schwankungsquadrat des Ortes die ber"uhmte Formel
\begin{equation}
\left\langle(\Delta x)^2\right\rangle=2Dt=\frac{kT}{3\pi\eta_0a}~t.
\end{equation}
Treffend sagt dazu Pais:
\begin{quote}
``\textit{Dieses Ergebnis ist jedesmal von neuem "uberraschend und kommt ganz unerwartet:
einige kleine -- aber im Vergleich zu Molek"ulen grosse -- Kugeln, eine Stoppuhr und ein
Mikroskop reichen aus, um Avogadro's Zahl zu bestimmen.}''
\end{quote}

All dies ist so bekannt, dass weitere Erl"auterungen unn"otig sind. Zur Abrundung hier noch
ein paar S"atze aus dem einleitenden Abschnitt, in welchem Einstein klar und deutlich
beschreibt worauf es ihm vor allem ankommt:
\begin{quote}
``\textit{In dieser Arbeit soll gezeigt werden, dass nach der molekularkinetischen Theorie
der W"arme in Fl"ussigkeiten suspendierte K"orper von mikroskopisch sichtbarer Gr"osse
infolge der Molekularbewegung der W"arme Bewegungen von solcher Gr"osse ausf"uhren m"ussen,
dass diese Bewegungen leicht mit dem Mikroskop nachgewiesen werden k"onnen. (...) Wenn sich
die hier zu behandelnde Bewegung samt den f"ur sie zu erwartenden Gesetzm"assigkeiten
wirklich beobachten l"asst, so ist die klassische Thermodynamik schon f"ur mikroskopisch
unterscheidbare R"aume nicht mehr als genau g"ultig anzusehen und es ist dann eine exakte
Bestimmung der wahren Atomgr"osse m"oglich. Erwiese sich umgekehrt die Voraussage dieser
Bewegung als unzutreffend, so w"are damit ein schwerwiegendes Argument gegen die
molekularkinetische Auffassung der W"arme gegeben.}''
\end{quote}

Sp"ater hat Einstein sehr wichtige Anwendungen der Brown'schen Bewegung in der
Quantentheorie gemacht. Mit deren Hilfe gelang es ihm, das Teilchenbild zu
vervollst"andigen. Ich komme darauf zur"uck.

\section{Kritische Opaleszenz}

Ein Brief von Einstein an Jakob Laub vom 27. August 1910 enth"alt den Passus:
\begin{quote}
``\textit{Ich schreibe gegenw"artig eine Arbeit "uber Opaleszenz von Gasen und
Fl"ussigkeiten. Quantitative Durchf"uhrung der Theorie von Smoluchowski. Mit dem
Prinzipiellen bin ich fertig. Die Theorie ist vollkommen streng.}''
\end{quote}
Darin leitet Einstein eine Formel f"ur die Streuung von Licht an Dichteschwankungen her.
Ein letztes mal findet er damit eine neue Methode, die Avogadro-Zahl zu bestimmen. An Laub
schreibt er dazu etwas sp"ater: ``Ich habe meine Freude daran''.  Dies war Einstein's
letzte gr"ossere Arbeit "uber \emph{klassische} statistische Physik. Die zugeh"origen
Messungen wurden kurz danach ausgef"uhrt.

Seit etwa 1874 war bekannt, dass die Streuung von Licht beim Durchgang durch ein Gas in der
N"ahe des kritischen Punktes wesentlich verst"arkt wird. Dieses Ph"anomen wurde 1908 durch
Smoluchowski auf grosse Dichteschwankungen zur"uckgef"uhrt. Darauf weist Einstein gleich zu
Beginn seiner Abhandlung hin:
\begin{quote}
``\textit{Smoluchowski hat in einer wichtigen theoretischen Arbeit gezeigt, dass die
Opaleszenz bei Fl"ussigkeiten in der N"ahe des kritischen Zustandes sowie die Opaleszenz
bei Fl"ussigkeitsgemischen in der N"ahe des kritischen Mischungsverh"altnisses und der
kritischen Temperatur vom Standpunkte der Molekulartheorie der W"arme in einfacher Weise
erkl"art werden kann.}''
\end{quote}
 Smoluchowski hat aber die Lichtstreuung an Dichteschwankungen nicht n"aher untersucht. Es
 war Einstein, der diese L"ucke ausf"ullte.

 Bevor er diese Aufgabe in Angriff nimmt, gibt er eine l"angere klare Darstellung von
 Schwankungserscheinungen auf der Basis des Boltzmann'schen Prinzips, die er dann auf
 Dichteschwankungen in Fl"ussigkeiten und Fl"ussigkeitsgemischen anwendet. Dieser
 er"offnende Teil ist ein Beitrag zur statistischen Thermodynamik, an den viele Forscher
 ankn"upften.

 Erst im vierten Abschnitt beginnt Einstein mit der ``Berechnung des von einem unendlich
 wenig inhomogenen absorptionsfreien Medium abgebeugten Lichtes'' im Rahmen der
 Maxwell'schen Theorie. Darin leitet er seine bekannte Formel f"ur den
 Extinktions- (Streu-) Koeffizienten her, die l"angst zum Vorlesungsstoff geworden ist. Falls der
 Brechungsindex $n$ nahe bei 1 ist, reduziert sich diese auf
 \begin{equation}
 \alpha(\omega)=\frac{1}{6\pi}\left(\frac{\omega}{c}\right)^4(n^2-1)^2\frac{kT}{-V(\partial p/\partial V)_T}
 \end{equation}
 ($\omega$= Kreisfrequenz des Lichtes). Einstein fand mit dieser Formel eine quantitative
 Verbindung zwischen Rayleigh-Streuung und kritischer Opaleszenz.

 Beim kritischen Punkt divergiert sein Ausdruck. Dann darf man, worauf Ornstein und Zernicke
 im Jahre 1915 hinwiesen, die Korrelationen der Dichtefluktuationen in verschiedenen
 Volumina nicht mehr vernachl"assigen. In diesem Sinne wurde Einstein's Arbeit "uber
 kritische Opaleszenz zum Ausgangspunkt von verschiedenen Forschungsrichtungen des
 zwanzigsten Jahrhunderts.

 \section{Einstein's Begr"undung der Planck'schen \\Verteilung, Impuls von Lichtquanten}

 Eine Art H"ohepunkt in Einstein's Bem"uhungen, so viel wie m"oglich aus der Planck'schen Verteilung
herauszuholen, war seine Arbeit ``Zur Quantentheorie der Strahlung'' von 1916. In dieser
f"uhrt er die bekannten statistischen Gesetze f"ur spontane und induzierte Emission sowie
f"ur Absorption ein, und findet damit eine Herleitung der Planck'schen Formel die zum
Lehrbuchstoff -- und zur theoretischen Grundlage des Lasers -- geworden ist. Mit einem
Korrespondenzargument im Rayleigh-Jeans Gebiet bekommt er zwischen den drei
`Einstein-Koeffizienten' zwei Beziehungen, die Dirac zehn Jahre sp"ater aus seiner
Quantentheorie der Strahlung ohne weitere Annahmen ableiten konnte. Einstein's Freude an
diesem Wurf kommt in einem Brief an Besso vom 11. August 1916 zum Ausdruck: ``Es ist mir
ein pr"achtiges Licht "uber die Absorption und Emission der Strahlung aufgegangen; es wird
Dich interessieren. Eine verbl"uffend einfache Ableitung der Planck'schen Formel, ich
m"ochte sagen \emph{die} Ableitung. Alles ganz quantisch.''

\subsection*{Nadelstrahlung}

Im zweiten Teil der Arbeit, den Einstein als den wichtigeren betrachtet, studiert er den
Impuls"ubertrag zwischen den Atomen und der Strahlung. Dabei benutzt er einmal mehr die
Theorie der Brown'schen Bewegung und zeigt mit einem sch"onen Argument, dass in jedem
Elementarprozess, insbesondere bei der \emph{spontanen} Emission, ein Impuls vom Betrag
$h\nu/c$ ausgetauscht wird. Dies steht nat"urlich in krassem Widerspruch zum klassischen
Bild der Strahlungsemission durch Kugelwellen. Der R"uckstoss eines Atoms bei der Emission
in eine zuf"allige Richtung wurde 1933 von R. Frisch experimentell best"atigt.

Da dieser Teil der Arbeit weniger bekannt ist, will ich Einstein's Argumentation etwas
ausf"uhren. F"ur die Untersuchung der Brown'schen Bewegung eines Atoms im thermodynamischen
Strahlungsfeld sind die folgenden "Uberlegungen wesentlich. Bewegt sich ein Atom mit der
Geschwindigkeit $v$, so "andert sich dessen Impuls $Mv$ in einer kurzen Zeit $\tau$ aus
zwei Gr"unden. Zum einen wirkt aufgrund von Absorptions- und induzierten Emissionsprozessen
eine systematische Reibungskraft $Rv$, welche zur Impuls"anderung $Rv\tau$ f"uhrt.
(Einstein's Berechnung von $R$ werden wir unten andeuten.) Diese Kraft allein w"urde das
Atom zur Ruhe bringen. Daneben bewirken aber die Schwankungen des Strahlungsfeldes eine
unregelm"assige "Anderung $\Delta$ in der Zeit $\tau$. Im Gleichgewicht gilt f"ur eine
eindimensionale Bewegung
\begin{equation}
\langle(Mv-Rv\tau+\Delta)^2\rangle=\langle(Mv)^2\rangle.
\end{equation}
Unter der Annahme $\langle v\cdot\Delta\rangle=0$ gibt dies
\begin{equation}
 \langle\Delta^2\rangle=2R\langle M^2v^2\rangle\tau=2RkT\tau.
\end{equation}

Nun wird die Rechnung zeigen, dass die systematische Reibungskraft $R$ durch folgenden
Ausdruck gegeben ist:
\begin{equation}
R=\frac{1}{3}\left(\frac{h\nu}{c}\right)^2\frac{1}{2kT}Z,
\end{equation}
wo $Z$ die Zahl der Elementarprozesse pro Sekunde bezeichnet. Nach (53) gilt also f"ur die
Fluktuation
\begin{equation}
\langle\Delta^2\rangle=\frac{1}{3}\left(\frac{h\nu}{c}\right)^2 Z\tau.
\end{equation}
Die Interpretation dieses Resultas liegt auf der Hand: Auch bei \textbf{spontaner} Emission
wird ein R"uckstoss $h\nu/c$ auf das Atom "ubertragen, denn dann ergibt sich f"ur $Z$
unabh"angige Elementarprozesse
\[ \langle\Delta^2\rangle=\left(\frac{h\nu}{c}\right)^2\langle\cos^2\theta\rangle Z\tau,
\] wo $\theta$ der R"uckstosswinkel ist, in "Ubereinstimmung mit (55).

Einstein kommentiert dieses Resultat in einem Brief an Besso vom 6. September 1916 kurz und
b"undig mit den Worten: ``Damit sind die Lichtquanten so gut wie gesichert'', und zwei
Jahre sp"ater bekr"aftigte er dies wiederum in einem Brief an Besso: ``Aber an der
Realit"at der Quanten zweifle ich nicht mehr, trotzdem ich mit dieser "Uberzeugung immer
noch ganz allein stehe. So wird es wohl bleiben, solange eine mathematische Theorie nicht
gelingt.''

Einstein widmete diese zentrale Arbeit dem Andenken von Alfred Kleiner; sie wurde zuerst in
den Mitteilungen der Physikalischen Gesellschaft Z"urich aus dem Jahre 1916 publiziert.

\subsubsection*{Zur Berechnung von $R$}

Das bewegte Atom sieht ein anisotropes Strahlungsfeld. Nun benutzt man, dass nach der
\emph{klassischen} Elektodynamik der Impuls eines Lichtstrahls dem Betrag nach gleich
dessen Energie dividiert durch $c$ ist. Mit Hilfe der Einstein'schen Koeffizienten $B_m^n$
kann man damit die Zahl der Absorptionen pro Zeiteinheit f"ur einen Atom"ubergang
$n\rightarrow m$ berechnen. Gleichzeitig ergibt sich dabei auch die zugeh"orige
Impuls"anderung etwa in der x-Richtung. Entsprechend erh"alt man den Beitrag von
induzierten Emissions"uberg"angen $m\rightarrow n$. Eine etwas l"angliche Rechnung, die bei
Einstein im Detail ausgef"uhrt wird, ergibt so das Resultat (54).

\section{Schlussbemerkungen}

Einstein geht in seinen autobiographischen Notizen, von ihm scherzhaft als Nekrolog
bezeichnet, nur kurz auf seine Anwendungen der klassischen Statistischen Mechanik ein. Die
Dissertation erw"ahnt er darin mit keinem Wort. Zu seinem Gesetz der Brown'schen Bewegung
meint er:
\begin{quote}
``\textit{Die "Ubereinstimmung dieser Betrachtung mit der Erfahrung zusammen mit der
Planck'schen Bestimmung der wahren Molek"ulgr"osse aus dem Strahlungsgesetz (f"ur hohe
Temperaturen) "uberzeugte die damals zahlreichen Skeptiker (Ostwald, Mach) von der
Realit"at der Atome. Die Abneigung dieser Forscher gegen die Atomtheorie ist ohne Zweifel
auf ihre positivistische philosophische Einstellung zur"uckzuf"uhren. Es ist dies ein
Beispiel daf"ur, dass selbst Forscher von k"uhnem Geist und von feinem Instinkt durch
philosophische Vorurteile f"ur die Interpretation von Tatsachen gehemmt werden k"onnen.}''
\end{quote}

Am Ende seines  ber"uhmten Buch ``Les Atomes'' schrieb Perrin r"uckblickend:
\begin{quote}
``\textit{Die Atomtheorie hat triumphiert. Ihre unl"angst noch zahlreichen Widersacher
verzichten, da sie endlich "uberzeugt sind, einer nach dem anderen auf die Einw"urfe,
welche lange Zeit berechtigt und ohne Zweifel n"utzlich waren.}''
\end{quote}

Einstein's guter Wert (49) fehlt "ubrigens in Perrin's Buch, das erstmals 1913 erschien. Da
Einstein erkannte, dass seine Dissertation noch nicht hinreichend bekannt war, machte er
1920 nochmals darauf aufmerksam, insbesondere um auf sein Erratum von 1911 hinzuweisen.

Mit seinem Standpunkt zur Strahlungstheorie, den er besonders deutlich in seiner Salzburger
Rede von 1909 deutlich machte, war Einstein seiner Zeit weit voraus. Bei diesem ersten
Auftritt vor einem grossen Kreis von Gelehrten w"ahlte er f"ur sein Referat den Titel
``\textit{"Uber die Entwicklung unserer Anschauungen "uber das Wesen und die Konstitution
der Strahlung}''. Bereits in den einleitenden Passagen sagt er prophetisch
\begin{quote}
``\textit{Deshalb ist es meine Meinung, dass die n"achste Phase der Entwicklung der
theoretischen Physik uns eine Theorie des Lichtes bringen wird, welche sich als eine Art
Verschmelzung von Undulations- und Emissionstheorie des Lichtes auffassen l"asst.}''
\end{quote}

W. Pauli bewertete in einem Aufsatz "uber Einstein's Beitr"age zur Quantentheorie diesen
Bericht ``als einen der Wendepunkte in der Entwicklung der theoretischen Physik''.

Es ist interessant was Einstein am Schluss der besprochenen Arbeit von 1916 "uber die Rolle
des Zufalls sagt:
\begin{quote}
``\textit{Die Schw"ache der Theorie liegt einerseits darin, dass sie uns dem Anschluss an
die Undulationstheorie nicht n"aher bringt, anderseits darin, dass sie Zeit und Richtung
der Elementarprozesse dem `Zufall' "uberl"asst; trotzdem hege ich das volle Vertrauen in
die Zuverl"assigkeit des eingeschlagenen Weges.}''
\end{quote}
Nach dieser richtungsweisenden Aussage erscheint es schwer verst"andlich, dass Einstein
sich str"aubte, den Durchbruch zur neuen Quantentheorie auf diesem Weg zu akzeptieren. Das
R"atsel der Lichtquanten liess ihn nie los. In einem Gespr"ach mit Res Jost erinnerte sich
Otto Stern, dass Einstein ihm einmal sagte: ``Ich habe hundertmal mehr "uber
Quantenprobleme nachgedacht, als "uber die allgemeine Relativit"atstheorie.''


\vspace{1cm}

\end{document}